\documentclass[aps,prl,twocolumn,superscriptaddress,longbibliography,10pt]{revtex4-1}
\usepackage{times}
\usepackage{amsmath}
\usepackage{bbold}
\usepackage{graphicx}
\usepackage{epstopdf}
\usepackage{upgreek}
\usepackage[english]{babel}
\usepackage{color}
\usepackage{bm}
\newcommand{\YIG}{\textsc{yig}}
\newcommand{\yig}{\textsc{yig}}
\usepackage{hyperref}
\hypersetup{
colorlinks=true,
linkcolor=blue,          
citecolor=blue,          
filecolor=magenta,       
urlcolor=cyan
}

\begin{document}

\title{Triple-resonant Brillouin light scattering in magneto-optical cavities}

\author{J. A. Haigh}
\email{jh877@cam.ac.uk}
\affiliation{Hitachi Cambridge Laboratory, Cambridge, CB3 0HE, UK}
\author{A. Nunnenkamp}
\affiliation{Cavendish Laboratory, University of Cambridge, Cambridge, CB3 0HE, UK}
\author{A. J. Ramsay}
\affiliation{Hitachi Cambridge Laboratory, Cambridge, CB3 0HE, UK}
\author{A. J. Ferguson}
\affiliation{Cavendish Laboratory, University of Cambridge, Cambridge, CB3 0HE, UK}

\date{\today}

\begin{abstract}
An enhancement in Brillouin light scattering (BLS) of optical photons with magnons is demonstrated in magneto-optical whispering gallery mode (WGM) resonators tuned to a triple-resonance point. This occurs when both the input and output optical modes are resonant with those of the whispering gallery resonator, with a separation given by the ferromagnetic resonance (FMR) frequency. The identification and excitation of specific optical modes allows us to gain a clear understanding of the mode-matching conditions. A selection rule due to wavevector matching leads to an intrinsic single-sideband excitation. Strong suppression of one sideband is essential for one-to-one frequency mapping in coherent optical-to-microwave conversion.
\end{abstract}

\maketitle

Extending microwave-optical transducers into a regime where inter-conversion between single optical and microwave photons is possible in a coherent manner \cite{regal_cavity_2011} is an important technological aim, as it would open up many avenues in, for example, implementing existing superconducting quantum devices \cite{devoret_superconducting_2013} in a wider quantum network \cite{kimble_quantum_2008}. Furthermore, frequency shifting of single photons would enable quantum optical devices to take advantage of wavelength division multiplexing. Strong progress towards these goals has been made in cavity optomechanics \cite{bochmann_nanomechanical_2013,andrews_bidirectional_2014,bagci_optical_2014,balram_coherent_2016}, and optimized electro-optic modulators \cite{strekalov_efficient_2009,chen_hybrid_2014,rueda_efficient_2016}.

Recently, microwave-optical inter-conversion has also been explored in a cavity opto-magnonic system \cite{osada_cavity_2016-1, zhang_optomagnonic_2015}, where magnetic Brillouin light scattering \cite{demokritov_brillouin_2001} has been reported in high $Q$ optical WGMs of a transparent magnetic sphere \cite{haigh_magneto-optical_2015}.  In this system, the collective excitations of the magnetic moment, magnons, play a role analogous to the phonons in a cavity optomechanics system \cite{brahms_spin_2010}. An important feature of this opto-magnonic system is the non-reciprocity of the BLS, where only one sideband has been observed \cite{osada_cavity_2016-1, zhang_optomagnonic_2015}.  A key requirement for a coherent transducer is a one-to-one mapping of the frequency components, and hence a strong suppression of one sideband. In contrast to an optomechanics system, due to conservation of angular momentum, optically induced creation and annihilation of magnons requires a change in optical polarization \cite{le_gall_theory_1971-1}. When combined with the geometric birefringence of a WGM resonator, this results in a non-reciprocal triple-resonance condition, where the optical pump and signal of opposite polarization are resonant with different cavity modes, whose frequency splitting is equal to the driven ferromagnetic resonance \cite{deych_resonant_2011}. Hence, side-band suppression is enforced by a selection-rule, rather than by detuning the pump laser from the optical cavity, as is usually the case in a cavity-optomechanics system.

\begin{figure}
\includegraphics[width=\columnwidth]{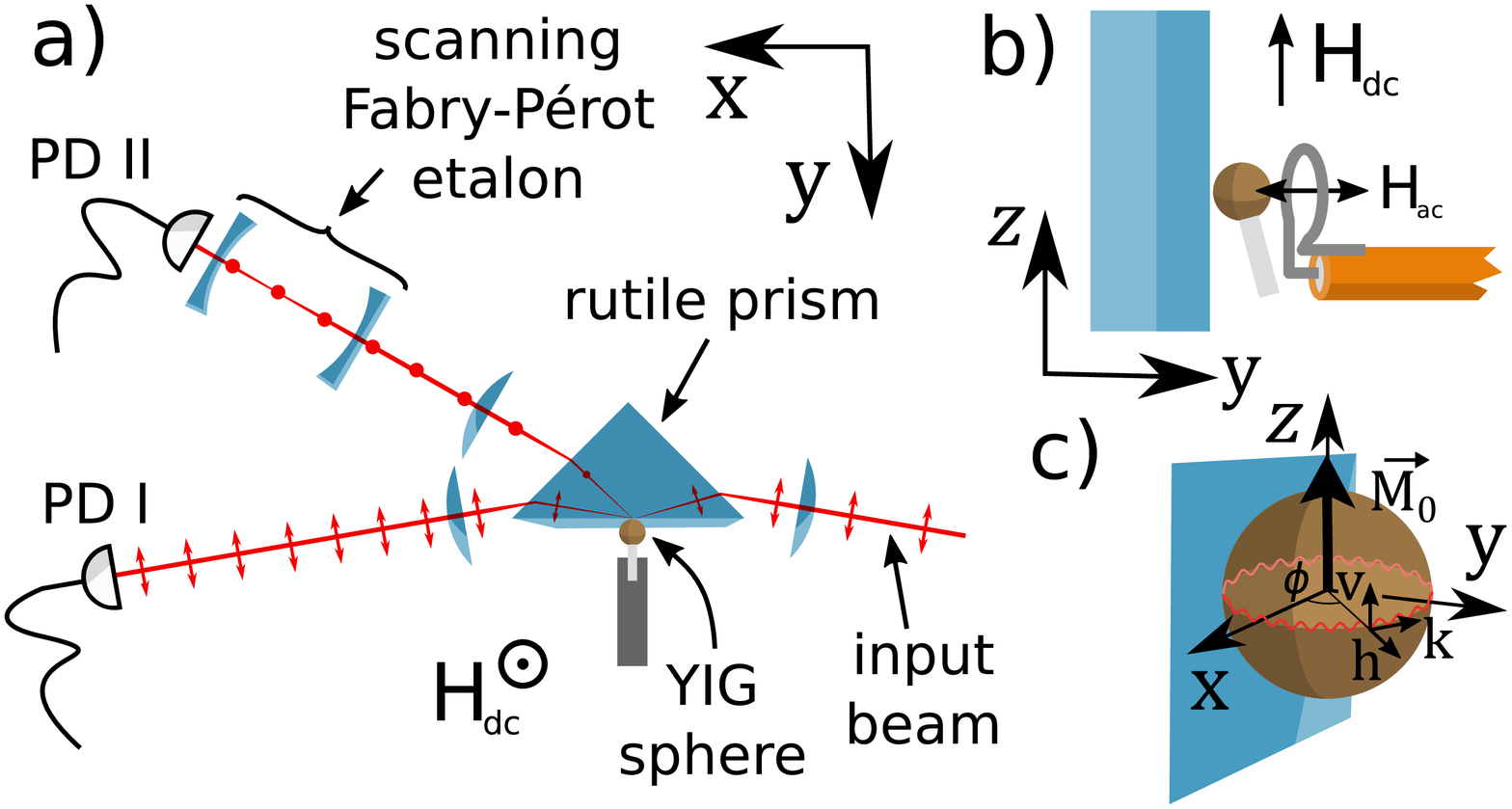}
\caption{(a) Top view of experimental setup. The scattered  light, with polarization orthogonal to the input beam, emitted at a different angle due to the birefringence of the rutile coupling prism, is spectroscopically analyzed with a scanning Fabry-P\'erot etalon. A dc magnetic field $H_\text{dc}$ is applied along the $z$-axis. (b) Microwave antenna to drive ferromagnetic resonance in the YIG sphere (side view). The microwave drive is provided by a vector network analyzer (VNA). The FMR modes are identified by measuring the microwave reflection coefficient $|S_{11}|$ as a function of frequency, after which the VNA is configured in continuous-wave mode to drive the resonance. (c) Coordinate systems used in the analysis.}
\label{fig1}
\end{figure}

In this Letter we show that the non-reciprocal triple-resonance condition between optical modes for pump and signal of the inter-conversion can be achieved with the precise mode identification allowed by prism coupling to the magnetic sphere. This is in contrast to previous measurements \cite{osada_cavity_2016-1, zhang_optomagnonic_2015}, where, due to the waveguide coupling used, the exact identification of the optical modes involved has been difficult, with the resonance condition being met accidentally \cite{zhang_optomagnonic_2015}.  For microwave driving of the uniform Kittel magnetization mode in the plane of the WGM, the polarization of the pump laser can be used to select the scattering direction via the fixed change in the azimuthal mode index. We identify that this selectivity arises from wavevector matching around the optical path of the pump and signal light-fields and the geometrical dependence of the magneto-optical coupling. Finally, measurements of the BLS intensity as a function of detuning from the triple-resonance condition show excellent agreement with a simple analytical model. Our experiments allow us to precisely characterize the resonant single-photon magneto-optical coupling strength \cite{liu_optomagnonics_2016, kusminskiy_coupled_2016}.

The experimental setup is shown in Fig.~\ref{fig1}(a). A prism coupler \cite{gorodetsky_optical_1999} is used to match the input angle, and therefore the wavevector, to the low order WGMs. The mode structure is probed by measuring the reflected intensity with same polarization as the input using a photodiode (PD~I), as the input laser wavelength is tuned. The light emitted from the cavity with opposite linear polarization to the input is emitted at a different angle due to the birefringence of the rutile prism. This polarization-scattered component is analyzed with a scanning Fabry-P\'erot etalon on an avalanche photodiode (PD~II). A microwave antenna (Fig.~\ref{fig1}(b)) is placed close to the YIG sphere to drive ferromagnetic resonance and the magnetic field from a permanent magnet (NdFeB) mounted on a stage is used to tune the FMR frequency. The setup can be switched to measure the same quantities for both linear polarizations of the input beam.

First, we identify the optical WGMs.  The dc magnetic field is fixed in the out-of-plane direction. Since there is no static component of the magnetization along the direction of propagation, mixing between linear polarized modes due to the Faraday effect is negligible \cite{haigh_magneto-optical_2015}. We therefore use the standard analytical forms of the WGM electric field distributions and resonant wavelengths \cite{gorodetsky_geometrical_2006, breunig_whispering_2013-1}, with two linearly polarized components perpendicular and parallel to the sphere surface. These modes, which we label horizontal $h$ and vertical $v$ \footnote{These modes are often labeled quasi-transverse magnetic (TM-$h$) and quasi-transverse electric (TE-$v$) \cite{osada_cavity_2016-1,zhang_optomagnonic_2015}.} (see Fig.~\ref{fig1}(c)), are split in energy due to the geometrical birefringence from the different surface boundary conditions for two electric field components.

\begin{figure}
\includegraphics[width=\columnwidth]{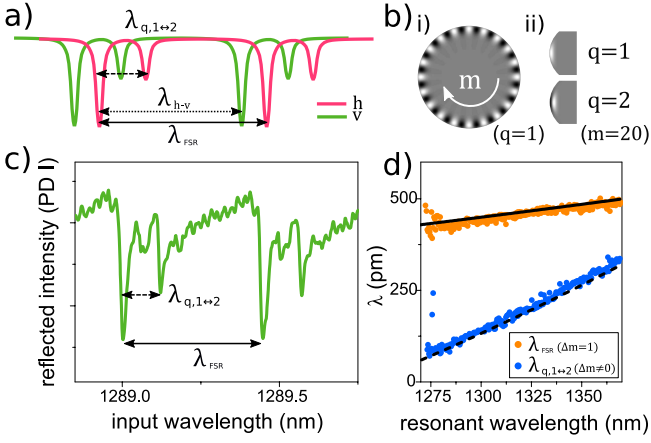}
\caption{Optical mode identification for an $r\approx250~\mu$m YIG sphere. (a) Schematic of mode families for radial indices $q=1,2$, and for $h$- and $v$-polarization. The free spectral range $\lambda_\textsc{fsr}$ is indicated by the black solid line. The $h$-$v$ splitting $\lambda_{h\mbox{-}v}$ is shown by the black dotted line, while the dashed line indicates the splitting between adjacent modes with $\Delta q = 1$. The azimuthal index $m$ determines the number of wavelengths around the circumference and radial index $q$ determines the number of radial nodes. (b) Representative plots of real part of the electric field (i) in the WGM plane for $m = 20$, $q=1$, and (ii) in cross-section for  $q = 1,2$. (c) Reflectance spectrum for $v$-polarized input. (d) The dispersion of the FSR $\lambda_\textsc{fsr}$ is used to identify the strongest mode family as $q=1$. The splitting between the different modes $\lambda_{q,1\leftrightarrow2}$ is used to identify the second as $q=2$. Solid and dashed lines show the calculated dispersions \cite{schunk_identifying_2014} fitted with small adjustments of the sphere radius.}
\label{fig2}
\end{figure}

The basic mode structure is shown schematically in Fig.~\ref{fig2}(a). The expected reflectance spectra for $h$ (pink) and $v$ (green) polarized modes are shown including modes with radial index $q=1,2$ for sets of modes with a difference $\Delta m=1$ in the azimuthal index $m$. The mode indices are defined in  Fig.~\ref{fig2}(b). The free spectral range is given by $\lambda_{\textsc{fsr}}=\lambda_0^2/2\pi r n_{\yig}$ to a good approximation in the relatively large spheres ($r\sim100~\mu$m, $m\sim1000$) which we study. In the same limit, the $h$-$v$ splitting is given by $\lambda_{h\mbox{-}v} = \lambda_{\textsc{fsr}}{\sqrt{n_{\YIG}^2 - 1}}/{n_{\YIG}}$ \cite{gorodetsky_geometrical_2006}. For YIG, with $n_{\yig}\approx 2.2$, $\lambda_{h\mbox{-}v}\approx 0.9\lambda_{\textsc{fsr}}$. Therefore the closest adjacent modes of opposite polarization are for different $m$ indices, separated by $m_v - m_h = 1$ and $\lambda^\text{eff}_{h\mbox{-}v}=0.1\lambda_{\textsc{fsr}}$.

Fig.~\ref{fig2}(c) presents a reflection spectrum for a $h$-polarized input. Two families of modes are observed. These are identified as $q=1$ and $q=2$ from comparison of the wavelength dispersion, shown in Fig.~\ref{fig2}(d), to the expected splitting. This demonstrates the highly selective excitation of the WGM, allowing clear identification of the matching conditions for enhanced wavelength conversion.

\begin{figure}
\includegraphics[width=\columnwidth]{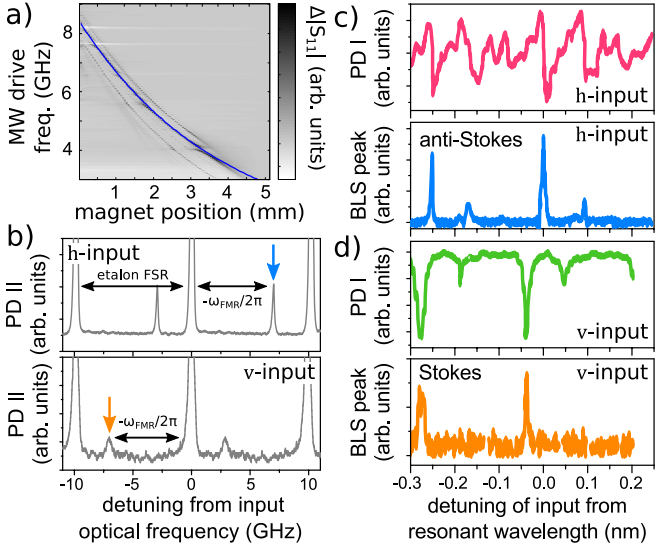}
\caption{(a) FMR of YIG sphere measured through microwave reflection coefficient $|S_{11}|$ of the antenna as a function of permanent magnet position. The blue dashed line shows the expected dependence of the uniform Kittel mode from the known dependence of magnetic field on distance from a cuboid magnet \cite{engel-herbert_calculation_2005} and the gyromagnetic ratio $\gamma=28$~GHz~T$^{-1}$. The magnetic field range is $H_\text{dc}\approx 100{\mbox{-}}320$~mT. The $Q$-factor of the magnetic mode is $Q_\textsc{fmr}\approx400$ (as this is due to Gilbert damping, the rate $\kappa_\textsc{fmr}\approx10{\mbox{-}}20$~MHz depends linearly on FMR frequency \cite{kambersky_spin-wave_1975}). (b) Measured spectra of emitted signal (PD~II) for $h$ (upper) an $v$ input polarization for $\omega_\textsc{fmr}/2\pi\approx7$~GHz. Orange and blue arrows label Stokes and anti-Stokes peaks, respectively. We can exclude the suppressed side-band down to the signal-to-noise ratio, maximum $\approx20$ (slightly different for the two input polarizations due to different experimental conditions). (c,d) Lower panels: maximum of BLS intensity as a function of input laser wavelength, for $h$ and $v$ input, respectively. Upper panels: Reflected optical intensity (PD~I), shown for comparison. The $x$-axes is detuning from the resonant wavelength of the $h$ polarized mode. For $v$ input measurements, this is set by the measured $h$-$v$ splitting $\lambda_{h\mbox{-}v}^\text{eff}$. For the optical modes $Q_v\approx2\times10^5$ and $Q_h\approx1\times10^5$ (dissipation rates $\kappa_v\approx1$~GHz, $\kappa_h\approx2$~GHz).}
\label{fig3}
\end{figure}

With the dc magnetic field in the out-of-plane direction $z$, we now introduce the microwave drive field in the in-plane $x$ direction. This drives ferromagnetic resonance (FMR), the precession of the magnetization about the static field. The magneto-static modes \cite{fletcher_ferrimagnetic_1959} of the YIG sphere can be identified by measuring the microwave reflection coefficient $S_{11}$ of the microwave antenna. The FMR spectrum as a function of permanent magnet position is shown in Fig.~\ref{fig3}(a) along with the expected Kittel mode frequency calculated from the position dependent magnetic field (blue line) \cite{engel-herbert_calculation_2005}. From this field dependence and the relative strength of the absorption, the uniform Kittel mode can be identified. During data collection the microwave drive tracks the FMR frequency to compensate for fluctuations in the dc magnetic field.

To achieve the triple-resonance condition, we use a sphere of radius $500~\mu$m, which has $\lambda^\text{eff}_{h\mbox{-}v}$ corresponding to $\omega_v-\omega_h\approx7$~GHz, and drive the FMR of the uniform Kittel mode close to that frequency. The cross-polarized emission of the cavity is spectrally analyzed using the etalon, and example data are shown in Fig.~\ref{fig3}(b). The top panel shows a measured spectrum for $h$ polarized input. There are two sets of peaks, each matching the 10~GHz FSR of the etalon. The largest is the elastic scattered light at the same wavelength as the input laser. The anti-Stokes signal is marked with a blue arrow and is higher in frequency by $\approx$7~GHz. There is no measurable Stokes peak for this input polarization for any input wavelength. The bottom panel shows a measured spectrum for $v$ polarized input. Here there is only a Stokes peak (orange arrow), lower in frequency by the microwave drive. In the following, we demonstrate that this asymmetry between the Stokes/anti-Stokes signal \cite{osada_cavity_2016-1, zhang_optomagnonic_2015}, different for the two input polarization, follows from a selection rule in the BLS process. The linewidth of the BLS peak is limited by the 200~MHz resolution of the etalon. We further note that when the magnetic field is reversed, the BLS is substantially reduced.

Figure~\ref{fig3}(c,d) compares the BLS peak amplitude as a function of detuning of the input laser from the resonance to the reflectivity spectra. The BLS is enhanced when the $h$($v$) polarized input laser is resonant with the $h$($v$) polarized, $q=1$, WGM.  

\begin{figure}
\includegraphics[width=\columnwidth]{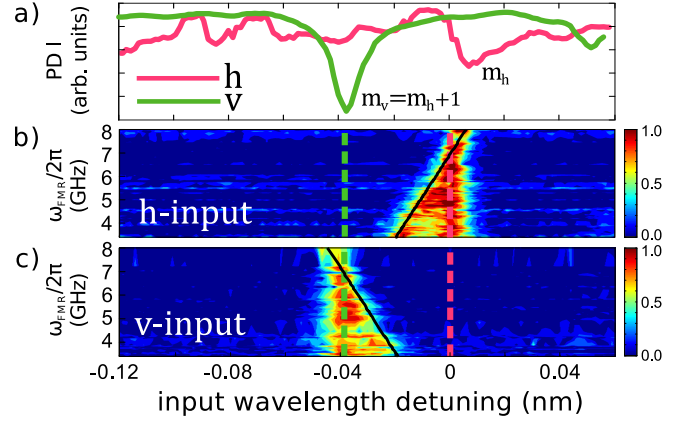}
\caption{BLS scattering amplitudes for different input linear polarizations ($h$: pink, $v$: green). (a) Reflected intensity (PD~I) for comparison. Azimuthal mode indices are labeled for clarity. The two curves are plotted on separate scales. (b) Color-plot of BLS intensity for $h$-input polarization as a function of input laser wavelength and FMR frequency. Each scan for fixed FMR frequency has been normalized to the peak amplitude for that scan. (c) As in (b), but for $v$-input polarization. Dashed lines in (b) and (c) indicate the resonant wavelengths for the two polarizations. The $x$-axis in all panels is in detuning from the resonant wavelength of the $h$ polarized mode. For $v$ input measurements, this is set by the measured $h$-$v$ splitting.}
\label{fig4}
\end{figure}

To explore the triple-resonance condition, the wavelength dependence of the BLS peak is measured as a function of the FMR frequency $\omega_\textsc{fmr}$. This is shown in Fig.~\ref{fig4} for (b)  $h$ and (c) $v$ input polarization. For $h$ ($v$) input, we only observe the Stokes (anti-Stokes) signal, and the color corresponds to the intensity of that signal. As the WGMs are sensitive to changes in sphere temperature with dissipated microwave power, the wavelength scans are aligned at the dip in reflected intensity (PD~I), and are normalized to the peak value for that FMR frequency in order to highlight the mode structure. An example of the reflected intensities (PD~I) for both input polarizations are shown for comparison in Fig.~\ref{fig4}(a), these are independent of the FMR frequency.

In Fig.~\ref{fig4}(b,~c) there are two maxima in the efficiency of the BLS process. The first peak is independent of the FMR frequency and is aligned with the WGM of the input polarization. This corresponds to a cavity enhancement of the input light field. For small FMR frequencies there is a second peak whose wavelength is linear in the FMR frequency. For $h$~($v$) polarized input, the black lines in Fig.~\ref{fig4}(b,~c) corresponding to $\omega_v-\omega_\textsc{fmr}$ and $\omega_h+\omega_\textsc{fmr}$ respectively are in reasonable agreement with the data. Hence, the second peak corresponds to a cavity enhancement of output light field, shifted by the FMR frequency.

By tuning the FMR frequency to match the $h$-$v$ splitting, we achieve the triple-resonance condition. This scattering is between modes of different azimuthal mode indices, $\Delta m = \pm1$. In fact, this is consistent with our expectation, as, in the frame of the light propagating around the mode, the in-plane driven magnetization rotates with respect to the direction of propagation. This means that the magnetic mode has an effective wavevector, and azimuthal integration of the electromagnetic energy leads to a selection rule $m_v - m_h = 1$ \footnote{see Supplementary Information}. It is this required change in mode index that allows the triple-resonance condition to be achieved for reasonable magnetic field strengths, as the FSR is approximately equal to the $h$-$v$ splitting so that the two modes with $m_v - m_h = 1$ are closely spaced in frequency. This is in contrast to previous work \cite{osada_cavity_2016-1, zhang_optomagnonic_2015}, which has suggested $\Delta m =0$, requiring substantially higher magnetic fields. We also note that in scattering the radial index $q$ is unchanged, $\Delta q=0$.

Furthermore, we can see that the Stokes/anti-Stokes asymmetry persists even detuned from the triple-resonance condition. This indicates that the asymmetry is not governed simply by the optical density of states. In fact, the selection rule $m_v - m_h = 1$ means that interaction Hamiltonian for the magnon mode $\hat{b}$ and two optical modes $\hat{a}_h^{}$, $\hat{a}_v^{}$, reduces to two terms \cite{Note2}, corresponding to the observed Stokes/anti-Stokes asymmetry, selected by the input polarization:
\begin{equation}
\hat H_\text{int} = \hbar G (\hat{b}_{}^{} \hat{a}_v^{\dagger} \hat{a}_h^{}+\hat{b}_{}^{\dagger} \hat{a}_h^{\dagger} \hat{a}_v^{}).
\end{equation}
Hence, the scattering process is non-reciprocal due to the wavevector matching around the WGM and azimuthal dependence of the magneto-optical coupling. From the known strength of the Faraday effect in YIG, we calculate the single-photon coupling rate $G=1$~Hz \cite{Note2}.

\begin{figure}
\includegraphics[width=\columnwidth]{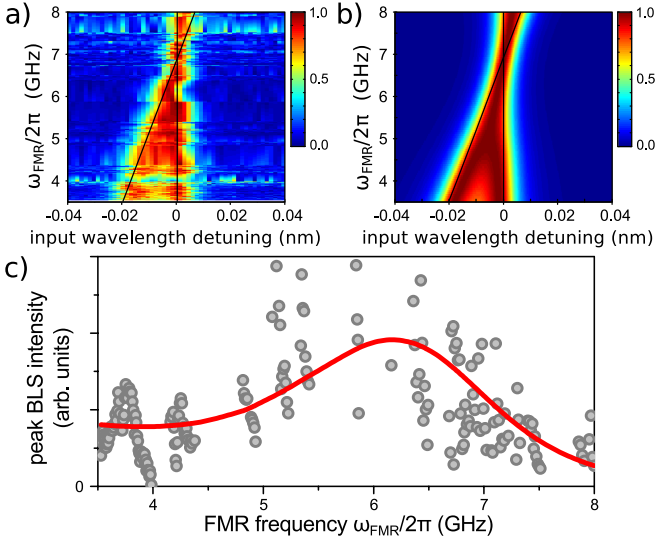}
\caption{Comparison between (a) experiment and (b) theory of BLS intensity as a function of input wavelength detuning and FMR frequency for $h$ input polarization. The black lines are the wavelengths corresponding to $\omega_h$ and $\omega_v - \omega_\textsc{fmr}$. Both experimental data and model are normalized to the peak value at each FMR frequency to allow better comparison of the mode structure. (c) Peak BLS efficiency as a function of FMR frequency. The red line is the expected trend given by the maximum of Eq.~(\ref{eq:amp}) for fixed $\omega_\textsc{fmr}$.}
\label{fig5}
\end{figure}

We can compare the measured data to a simple analytical model based on these three modes \cite{Note2}. The amplitude of the scattered field as a function of the detuning from the triple-resonance condition $\omega_\textsc{fmr}-\omega_v+\omega_L$ and of the $h$-polarized input frequency $\omega_h-\omega_L$ is
\begin{align}
|\langle & \hat{a}_{v,\textrm{out}} \rangle|^2 =\label{eq:amp} \\ 
&\frac{4 G^2  |\bar{a}_{h,\textrm{in}}|^2 |\bar{b}_\text{in}|^2\kappa_v\kappa_h/\kappa_\textsc{fmr}}{\left[\frac{\kappa_h^2}{4}+(\omega_h-\omega_L)^2\right] \left[\frac{\kappa_v^2}{4}+(\omega_\textsc{fmr}-\omega_v+\omega_L)^2\right]}. \nonumber
\end{align}
This is the product of two Lorentzians, corresponding to resonant enhancement of the input and output fields respectively. All the parameters are known from independent measurements, so that we can plot this expression in Fig.~\ref{fig5}(b), with excellent agreement with the data plotted alongside (Fig.~\ref{fig5}(a)). 

Finally, we plot the maximum BLS amplitude for each FMR frequency in Fig.~\ref{fig5}(c). The variation in the data is due to changes in the microwave power transmitted to the YIG sphere at different frequencies. The red line is the expected value given by Eq.~(\ref{eq:amp}), vertically scaled to match the data, with good agreement in the general trend.

To summarize, we have demonstrated the tuning of a cavity magneto-optical system to a triple-resonance condition for enhanced Brillouin light scattering. A selection rule $\Delta m = \pm1$ in the azimuthal index of the optical mode arises due to wavevector matching around the optical path of the WGM. Due to conservation of total angular momentum, a change in the optical orbital angular momentum of $\Delta m = \pm1$, results in the annihilation/creation of one magnon, and up/down-conversion of the light, respectively. The modes closest to energy-matching conditions have $m_v - m_h = 1$, and hence the polarization of the input laser selects either a Stokes or Anti-Stokes frequency conversion. Since the asymmetry of the BLS arises from a selection rule, a strong asymmetry can also be observed away from cavity resonance. This mechanism has similarities to BLS between two optical modes in optomechanics \cite{bahl_observation_2012-1}. We further note that non-transverse components of the optical modes \cite{junge_strong_2013} are not included in our model, and are therefore not needed to explain the asymmetry in the BLS \cite{osada_cavity_2016-1, zhang_optomagnonic_2015}.

Asymmetries in magnon BLS have been reported previously due to other mechanisms. Localization of surface magnon modes with a given chirality \cite{grunberg_light_1977} is not relevant here, as we study the uniform magnetic mode, and spin-spin correlations between different components introduced by the demagnetizing field are only relevant in a thin film geometry \cite{camley_stokeschar22anti-stokes_1982}. It is possible that interference between the first- (Faraday) and second-order (Voigt) magneto-optical effects \cite{le_gall_theory_1971-1, wettling_relation_1975} may result in minor corrections to the differing amplitudes.

Although the single-photon coupling rate is significantly smaller than the linewidths of the optical and magnetic modes, the scaling of the coupling with the magnetic mode volume suggests that interesting regimes could be achieved with smaller devices. Optimizing the $Q$-factor mode volume ratio \cite{akahane_high-q_2003} could be achieved with photonic-crystal defect cavities \cite{yoshie_vacuum_2004}, or plasmonic resonances, which have been shown to enhance static magneto-optical parameters \cite{safarov_magneto-optical_1994, jain_surface_2009, temnov_active_2010}. In addition, for the current system, cooling to low temperatures \cite{treussart_evidence_1998} should improve the $Q$-factors of all modes significantly. This may open up the possibility of spontaneous Brillouin cooling \cite{bahl_observation_2012-1} or lasing \cite{spillane_ultralow-threshold_2002} of the magnetic mode in this system.

\textit{Acknowledgments.} This project was partly funded by EPSRC under EP/M50693X/1. A.~J.~F. is supported by ERC grant 648613 and a Hitachi Research Fellowship. A.~N. holds a University Research Fellowship from the Royal Society and acknowledges support from the Winton Programme for the Physics of Sustainability.


\bibliography{bibliography}

\onecolumngrid
\clearpage
\setcounter{equation}{0}
\setcounter{figure}{0}
\setcounter{table}{0}

\makeatletter
\renewcommand{\theequation}{S\arabic{equation}}
\renewcommand{\thefigure}{S\arabic{figure}}

\begin{center}
\textbf{\large Supplemental Materials: Triple-resonant Brillouin light scattering in magneto-optical cavities}	
\end{center}
\setcounter{page}{1}

\section{Coupling constant and selection rules}
To derive the Hamiltonian describing magneto-optical coupling we start from the time-averaged electromagnetic energy
\begin{equation}
u =  \frac{1}{4} \int{dV} E_i^{*} \varepsilon_{ij} E_j^{}
\label{eq:}
\end{equation}
with hermitian dielectric tensor $\varepsilon_{ij} = \varepsilon_0\left(\varepsilon^{r}_i \delta_{ij} -  i \epsilon_{ijn} f M_n^{}\right)$, Faraday constant $f$, anisotropic dielectric constant $\varepsilon_0\varepsilon^{r}_i$, Levi-Civita symbol $\epsilon_{ijk}$, and magnetization component $M_n$. The electric fields of the WGM modes in dielectric spheres have approximate analytical forms \cite{schunk_identifying_2014, gorodetsky_geometrical_2006}. Here, we are interested  in a single pair of linearly polarized modes with polarization vector $\vec{v}$ perpendicular to the WGM plane and $\vec{h}$ perpendicular to the sphere surface, see Fig.~1(c).
In a spherical coordinate system with axis along $\vec{h}$, $\vec{v}$, and the direction of propagation $\vec{k}=(\vec{h}\times\vec{v})$, we can write
\begin{equation}
\vec{E} = E_h^{} \vec{h} + E_v^{} \vec{v}.
\label{eq:}
\end{equation}
The component of the magnetization which enters the energy is that along $\vec{k}$, i.e.~$M_{k} = -M_x \sin \phi + M_y \cos \phi$ for a counter-clockwise direction of propagation. Substituting $M_{\boldsymbol{\pm}} = M_x \pm i M_y$ we obtain $M_k=-i\frac{1}{2}\left(M_{\boldsymbol{+}} e^{ -i\phi}  - M_{\boldsymbol{-}} e^{ i\phi}\right)$.
The coupling constant can be found from the magneto-optical coupling part of $u$,
\begin{align}
u_\text{int} &= - \frac{i}{4}\varepsilon_0 f \int{dV   \left(M_k^{} E_h^* E_v^{} - M_k^{} E_v^* E_h^{}\right)}\\
&= - \frac{1}{8}\varepsilon_0 f \int dV   \left( e^{ - i\phi} M_{\boldsymbol{+}}^{} E_h^* E_v^{} - e^{ i\phi} M_{\boldsymbol{-}}^{} E_h^* E_v^{}\right.\nonumber \\
& \hspace{2.2cm} \left. - e^{ - i\phi} M_{\boldsymbol{+}}^{} E_v^* E_h^{} + e^{ i\phi} M_{\boldsymbol{-}}^{} E_v^* E_h^{} \right).\nonumber
\label{eq:}
\end{align}
The electric field is quantized \cite{kusminskiy_coupled_2016} by substituting
\begin{equation}
E_{h,v}\rightarrow\hat{E}^{{\boldsymbol{+}}}_{h,v} = i \sqrt{\frac{\hbar \omega_{h,v}}{2 \varepsilon_0 n_\YIG^2 V_\textsc{wgm}}} \mathcal{F} (r,\theta) \frac{1}{\sqrt{2 \pi}} e^{- i m_{h,v} \phi} \hat{a}_{h,v}^{}
\label{eq:}
\end{equation}
where $m_{h,v}$ is the azimuthal mode index, defined positive for counter-clockwise propagation, and $\mathcal{F}(r,\theta)$ is the mode function, normalized such that $\int{\mathcal{F}}dV = V_\textsc{wgm}$ approximately equal for the two modes. We leave the azimuthal part outside the mode function to emphasize a selection rule which will become apparent, and replace the magnetization components with the raising and lowering operators for the total spin $M_+ \rightarrow (M_0/S_0)\hat{S}_+$, where $M_0$ and $S_0$ are the total magnetization and spin, respectively. This gives
\begin{align}
\hat{H}_\text{int} = - \frac{\varepsilon_0 f}{8 } \frac{\hbar \omega}{2 \varepsilon_0 n_\YIG^2} &\frac{M_0}{S_0} \frac{1}{2\pi} \int d\phi   \\
													 \times & \left[    e^{-i\phi(1-m_h+m_v)} \hat{S}_{\boldsymbol{+}}^{} \hat{a}_h^{\dagger} \hat{a}_v^{} -  e^{i\phi(1+m_h-m_v)} \hat{S}_{\boldsymbol{-}}^{} \hat{a}_h^{\dagger} \hat{a}_v^{} \right.\nonumber \\ 
												  - &  \left. e^{-i\phi(1+m_h-m_v)} \hat{S}_{\boldsymbol{+}}^{} \hat{a}_v^{\dagger} \hat{a}_h^{} +  e^{i\phi(1-m_h+m_v)} \hat{S}_{\boldsymbol{-}}^{} \hat{a}_v^{\dagger} \hat{a}_h^{} \right], \nonumber
\label{eq:}
\end{align}
where we have used the approximation $\omega = \omega_v\approx\omega_h$. 

In the experiment, the optical cavity modes of interest are separated by $\approx 7$~GHz, of the order of the FMR frequency, and have $m_{v}-m_{h}=1$. This satisfies the selection rule given by the azimuthal integration for second and third terms. The first and fourth terms would be satisfied by $m_{v}-m_{h}=-1$, but the frequency separation of these modes is $\approx90$~GHz, such that the triple-resonance condition is far from being met. This results in the measured Stokes/anti-Stokes asymmetry. We note that for the opposite direction of optical propagation the mode indices will change sign $m_h\rightarrow-m_h$, such that the opposite terms would survive. For the situation of interest, we can write down the interaction Hamiltonian
\begin{align}
\hat H_\text{int} =  \hbar g (\hat{S}_{\boldsymbol{-}}^{} \hat{a}_h^{\dagger} \hat{a}_v^{} + \hat{S}_{\boldsymbol{+}}^{} \hat{a}_v^{\dagger} \hat{a}_h^{}),
\end{align}
with coupling constant
\begin{equation}
g =  \frac{\varepsilon_0 f M_0}{8 \hbar S_0} \frac{\hbar \omega}{2 \varepsilon_0 n_\YIG^2}.
\label{eq:}
\end{equation}
The expression can be simplified by putting Faraday constant in terms of Verdet constant $f = (2 c n_\YIG / \omega M_0) \mathcal{V}$, using $S_0 = N_\text{spins}/2$,
\begin{equation}
g =  \frac{1}{4 N_\text{spins}} \frac{c}{n_\YIG} \mathcal{V},
\label{eq:}
\end{equation}
where $N_\text{spins}$ is the number of spins in the sphere. Using the Holstein-Primakoff approximation the interaction Hamiltonian can also be written in terms of magnon creation and annihilation operators $\hat{b}^{\dagger}$, $\hat{b}$, where the association with the raising and lowering operators $\hat{S}_{\boldsymbol{\pm}}$ will depend on the equilibrium direction of the magnetization. For dc magnetic field in the $+z$ direction, the substitution is $\hat{S}_{\boldsymbol{-}} = \sqrt{2 S_0} \hat{b}_{}^\dagger $, which results in Eq.~(1) of the main text
\begin{equation}
\hat{H}_\text{int} = \hbar G (\hat{b}_{}^{} \hat{a}_v^{\dagger} \hat{a}_h^{} + \hat{b}_{}^{\dagger} \hat{a}_h^{\dagger} \hat{a}_v^{}),
\label{eq:}
\end{equation}
with the single-photon coupling strength
\begin{equation}
G = \sqrt{N_\text{spins}} g =  \frac{ \mathcal{V}  c'}{4}\sqrt{ \frac{1}{ N_\text{spins}} },
\label{eq:}
\end{equation}
where $c' = c/n_\YIG$ is the speed of light in YIG \cite{osada_cavity_2016-1}. The terms in $\hat{H}_\text{int}$ then satisfy energy conservation, as $\omega_v<\omega_h$, e.g. for $b_{}^{\dagger} \hat{a}_h^{\dagger} \hat{a}_v^{}$ we have $\omega_\textsc{fmr} = \omega_h-\omega_v$. This would not be the case for the opposite magnetic field direction, $\hat{b}^\dagger\rightarrow \hat{b}$. We observe this in experiment; for opposite direction of the magnetic field the Stokes/anti-Stokes symmetry is reversed but the amplitude is severely reduced. Note that equivalently for the opposite direction of optical propagation, the opposite magnetic field direction would be required.

For a 1~mm sphere we estimate $G/ 2 \pi \approx 1$~Hz, from the number density of spins $n_\text{spin}=2.1\times10^{28}$~m$^{-3}$, $V_\text{sphere} = 4.2\times10^{-10}$~m$^{3}$ ($N_\text{spins}=1\times10^{19}$), $\mathcal{V}=3.77$~rad~cm$^{-1}$ \cite{osada_cavity_2016-1}, and $n_\YIG=2.2$.


\section{Dynamical model}

The derivation of Eq.~(2) in the main text is as follows. Our system is described by the Hamiltonian
\begin{equation}
\hat{H} = \hbar\omega_h \hat{a}^\dagger_h \hat{a}_h^{} + \hbar\omega_v \hat{a}^\dagger_v \hat{a}_v^{} + \hbar \omega_\textsc{fmr} \hat{b}^\dagger_{} \hat{b} + \hbar G (\hat{b}^{}_{}\hat{a}^\dagger_v \hat{a}_h^{}+\hat{b}^{\dagger}_{}\hat{a}^\dagger_h \hat{a}^{}_v)
\end{equation}
In the frame of the laser drive, separating mean and fluctuations, we have $\hat{a}_{h} = e^{-i \omega_L t} (\bar{a}_{h} + \hat{d}_{h})$ and $\hat{a}_{v} = e^{-i \omega_L t} (\bar{a}_{v} + \hat{d}_{v})$. Driving the $h$-polarized mode with drive strength $\bar{a}_\textrm{h,in}$ leads for small coupling $G$ to $\bar{a}_v = 0$ and
\begin{equation}
\bar{a}_h = \frac{\sqrt{\kappa_h}\bar{a}_\textrm{h,in}}{\frac{\kappa_h}{2}-i(\omega_h-\omega_L)}.
\end{equation}
For small coupling $G$ we can linearize the Hamiltonian
\begin{equation}
\hat{H}'= \hbar(\omega_h-\omega_L) \hat{d}^\dagger_h \hat{d}_h^{} + \hbar(\omega_v-\omega_L) \hat{d}^\dagger_v \hat{d}_v^{} + \hbar\omega_\textsc{fmr} \hat{b}^\dagger \hat{b} + \hbar G |\bar{a}_h| (\hat{b}^\dagger \hat{d}_v^{} + \hat{b} \hat{d}^\dagger_v).
\end{equation}

From this we derive quantum Langevin equations assuming standard linear damping for both optical modes $\kappa_h$, $\kappa_v$ as well as the magnon mode $\kappa_\textsc{fmr}$ (this damping rate is linear in $\omega_\textsc{fmr}$, corresponding to the Gilbert damping in the Landau-Lifshitz equation \cite{kambersky_spin-wave_1975})
\begin{align}
\dot{\hat{b}} = &-i \omega_\textsc{fmr} \hat{b} - \frac{\kappa_\textsc{fmr}}{2}\hat{b} + \sqrt{\kappa_\textsc{fmr}} \hat{b}_\textrm{in} - i \bar G \hat{d}^{}_v\\
\dot{\hat{d}}_v = &-i (\omega_v-\omega_L)\hat{d}_v - \frac{\kappa_v}{2}\hat{d}_v + \sqrt{\kappa_v} \hat{d}_\textrm{v,in} - i \bar G \hat{b}^{}
\end{align}
with the driving-enhanced coupling constant $\bar G = G |\bar{a}_h|$.

In the Fourier domain we obtain
\begin{align}
\hat{b}(\omega) = &\frac{+ \sqrt{\kappa_\textsc{fmr}} \hat{b}_\textrm{in} - i \bar G \hat{d}^{}_v}{\frac{\kappa_\textsc{fmr}}{2} -i (\omega-\omega_\textsc{fmr})}\\
\hat{d}_v(\omega) = &\frac{+ \sqrt{\kappa_v} \hat{d}_\textrm{v,in} - i \bar G \hat{b}^{}}{\frac{\kappa_v}{2} -i [\omega-(\omega_v-\omega_L)]}.
\end{align}

In the experiment we drive FMR resonantly, $\langle \hat{b}_\text{in}(\omega) \rangle = \bar{b}_\text{in} 2\pi \delta(\omega-\omega_\textsc{fmr})$, so the coherent response to second order in $G$ is
\begin{equation}
|\langle \hat{d}_v \rangle|^2
= \frac{4 G^2 |\bar{a}_{h,\text{in}}|^2 |\bar{b}_\text{in}|^2 \kappa_h /\kappa_\textsc{fmr}}{\left[\frac{\kappa_h^2}{4}+(\omega_h-\omega_L)^2\right] \left[\frac{\kappa_v^2}{4}+(\omega_\textsc{fmr}-\omega_v+\omega_L)^2\right]}
\end{equation} 
where we integrated over the frequency $\omega$ as the resolution of the spectrometer is of the order of the cavity linewidth $\kappa_\textrm{h}$ and much lower than the FMR linewidth $\kappa_\textsc{fmr}$.

With the input-output relation $\hat{a}_{v,\text{out}}=\hat{a}_{v,\text{in}}-\sqrt{\kappa_v} \hat{a}_v$ we finally obtain Eq.~(2) of the main text as
\begin{equation}
|\langle \hat{a}_{v,\text{out}} \rangle|^2
= \frac{4 G^2 |\bar{a}_{h,\text{in}}|^2 |\bar{b}_\text{in}|^2 \kappa_v \kappa_h/\kappa_\textsc{fmr}}{\left[\frac{\kappa_h^2}{4}+(\omega_h-\omega_L)^2\right] \left[\frac{\kappa_v^2}{4}+(\omega_\textsc{fmr}-\omega_v+\omega_L)^2\right]}.
\end{equation}

\end{document}